\documentclass[journal]{IEEEtran}
\usepackage{balance}
\usepackage{graphicx}
\usepackage{bm}
\usepackage{amsmath}
\usepackage{amsfonts}
\usepackage{amsthm}
\usepackage{amssymb}
\usepackage{color}
\usepackage{mathrsfs}
\usepackage{upgreek}
\usepackage[utf8]{inputenc}
\usepackage{tikz}
\usetikzlibrary{arrows.meta,calc}
\usepackage{cite}
\usepackage{booktabs}
\usepackage{multirow}
\usepackage{amsmath}
\usepackage{amsthm}
\usepackage{algorithm}
\usepackage{algorithmic}
\usepackage{tabularx}
\usepackage{threeparttable}
\ifCLASSOPTIONcompsoc
 \usepackage[caption=false,font=normalsize,labelfont=sf,textfont=sf]{subfig}
\else
 \usepackage[caption=false,font=footnotesize]{subfig}
\fi

\usepackage{hyperref}					
\hypersetup{colorlinks,
	linkcolor=blue,%
	anchorcolor=blue,
    citecolor=blue}

\begin{document}
\title{{{Relativistic Cram\'er-Rao Bound Scaling for \\Device-Based and Device-Free Sensing}}
}
\author{
	{
	Fan Liu,~\IEEEmembership{Senior Member,~IEEE}, Yifeng Xiong,~\IEEEmembership{Member,~IEEE}, Weijie Yuan,~\IEEEmembership{Senior Member,~IEEE}, \\Yuanhao Cui,~\IEEEmembership{Member,~IEEE},~Jie Yang,~\IEEEmembership{Member,~IEEE},~and~Shi Jin,~\IEEEmembership{Fellow,~IEEE}
	} 
\thanks{This work was supported in part by the National Natural Science Foundation of China (NSFC) under Grant 62522107, and in part by the Fundamental Research Funds for the Central Universities under Grant 2242026RCB0200. (\textit{Corresponding authors: Yifeng Xiong; Shi Jin.})}
\thanks{F. Liu and S. Jin are with the National Mobile Communications Research Laboratory, Southeast University, Nanjing 210096, China (e-mail: fan.liu@seu.edu.cn, jinshi@seu.edu.cn).}
\thanks{Y. Xiong is with the School of Information and Electronic Engineering, Beijing University of Posts and Telecommunications, Beijing 100876, China. (e-mail: yifengxiong@bupt.edu.cn)}
\thanks{W. Yuan is with the School of Automation and Intelligent Manufacturing, Southern University of Science
and Technology, Shenzhen 518055, China. (email: yuanwj@sustech.edu.cn).}
\thanks{Y. Cui is with the School of Information and Electronic Engineering, Beijing University of Posts and Telecommunications, Beijing 100876, China (e-mail: yuanhao.cui@bupt.edu.cn).}
\thanks{J. Yang is with the Frontiers Science Center for Mobile Information Communication and Security, Southeast University, Nanjing 210096, China, and also with the Key Laboratory of Measurement and Control of Complex Systems of Engineering, Ministry of Education, Southeast University, Nanjing 210096, China (email: yangjie@seu.edu.cn).}
}
\maketitle

\begin{abstract}
This letter investigates range and velocity estimation under relativistic motion for device-based (DB) and device-free (DF) sensing. By deriving the exact time-scaling and time-shift relations induced by one-way and two-way propagation, both sensing modes are cast into a unified affine signal model. Closed-form Cramér--Rao bounds (CRBs) are obtained as explicit functions of normalized velocity, root-mean-squared (RMS) bandwidth, and RMS duration. The bounds recover the classical low-speed results but exhibit distinct velocity scaling in the ultrarelativistic regime. For rapidly receding motion, the range CRB diverges while the velocity CRB vanishes. For rapidly approaching motion, both CRBs vanish. The DB and DF modes further exhibit different asymptotic orders in the two directions, showing that relativistic motion changes not only the signal model but also the fundamental scaling laws governing sensing accuracy.
\end{abstract}
\begin{IEEEkeywords}
Wireless sensing, special relativity, estimation theory, signal processing, speed of light.
\end{IEEEkeywords}

\section{Introduction}
\IEEEPARstart{I}{n} most wireless sensing applications, relative velocities are so small compared with the speed of light that spacetime appears effectively Newtonian: clocks may be treated as sharing an absolute time, and motion is well described through a propagation delay and a Doppler frequency shift. This approximation is remarkably successful, but it hides the relativistic time relation that actually governs electromagnetic propagation \cite{Einstein1905}. Under the usual narrowband assumption on the transmitted signal $s(t)$ and the low-speed assumption on the relative motion, the received signal is modeled as
\begin{equation}
y(t)=\alpha s(t-\tau)e^{j2\pi \nu t}+z(t),
\label{eq:classical_signal}
\end{equation}
where $z(t)$ is the additive white Gaussian noise (AWGN), and $\tau$ and $\nu$ are determined by the range and radial velocity, respectively. This model leads to the well-known result that ranging accuracy is governed by signal bandwidth, whereas velocity estimation accuracy is governed by signal duration \cite{Friedlander1984,Dogandzic2001}. As the relative velocity becomes comparable with the speed of light, however, the Newtonian approximation breaks down. Motion no longer appears merely as a delay and a frequency shift. It compresses or stretches the waveform time axis itself. The sensing problem must then be formulated through the relativistic relation between transmitted and received time. This raises a basic question: \textit{What are the fundamental limits of range and velocity estimation in this regime?}

Relativistic effects in electromagnetic measurements have long been studied in physics and astronomy. The relativistic Doppler relation is routinely used to interpret spectral shifts and infer radial velocities, while precision astronomical observations also account for observer motion and reference time scales \cite{Lindegren2003,Wright2014}. Related issues arise in satellite and deep space tracking, where range and Doppler observables are obtained from relativistic light propagation and clock relations \cite{Harkins1979,Turyshev2013,Hees2014}. Relativistic deformation of pulsed electromagnetic waveforms has also been studied directly in the time domain \cite{deHoop2009}. From a signal processing perspective, relative motion can be described through the mapping between transmission and reception time. At low velocity and under the narrowband approximation, this mapping reduces to a Doppler frequency shift. For wideband signals, the time deformation must instead be retained as a Doppler stretch. This has motivated studies of wideband matched filtering and ambiguity functions \cite{Kelly1965,Rihaczek1967,Altes1973,Sibul1981,White1992,Weiss1994,Borden2006}, as well as joint delay and stretch estimation and waveform design \cite{Jin1995,Giunta1998,Niu1999,Das2020}. In most such work, however, the stretch factor is introduced through a classical moving target model rather than derived from the relativistic timing and propagation relations of the sensing system.

The relevant time relation also depends on how the sensing signal is generated and observed. A common classification in radio sensing distinguishes two basic cases \cite{Liu2022FundamentalLimits}. In DB sensing, a moving device actively transmits a signal to a receiver through one-way propagation, so transmission follows the clock of the device while reception is referenced to the receiver clock. In contrast, monostatic DF sensing uses a transceiver to illuminate a passive moving target and receive its reflection. Transmission and reception therefore share the same clock, with the signal undergoing two-way propagation through the moving target. These different timing structures become significant at relativistic velocities. It is therefore not obvious how the two sensing modes compare under the same received signal-to-noise ratio (SNR), or how their range and velocity estimation limits behave as the object velocity approaches the speed of light.

In this letter, we characterize the range and velocity estimation limits of DB and DF sensing under relativistic motion. By commencing with the exact timing relations imposed by special relativity and electromagnetic propagation, we derive the received signal models and closed-form CRBs as explicit functions of the normalized velocity, RMS bandwidth, and RMS duration. The bounds recover the classical low-speed results but exhibit distinct velocity scaling near the speed of light. For rapidly receding motion, the range Fisher information vanishes while the velocity information grows without bound. For rapidly approaching motion, both range and velocity CRBs vanish. DB and DF sensing further exhibit different asymptotic orders in the two directions, and under equal received SNR their relative performance can even reverse at sufficiently high velocities. These results reveal velocity-dependent estimation limits of DB and DF sensing that are absent from the classical delay--Doppler model. More broadly, they show that relativistic motion changes not only the signal model, but also the fundamental scaling laws that govern sensing accuracy.


\begin{figure}[!t]
	\centering
	\includegraphics[width = 0.95\columnwidth]{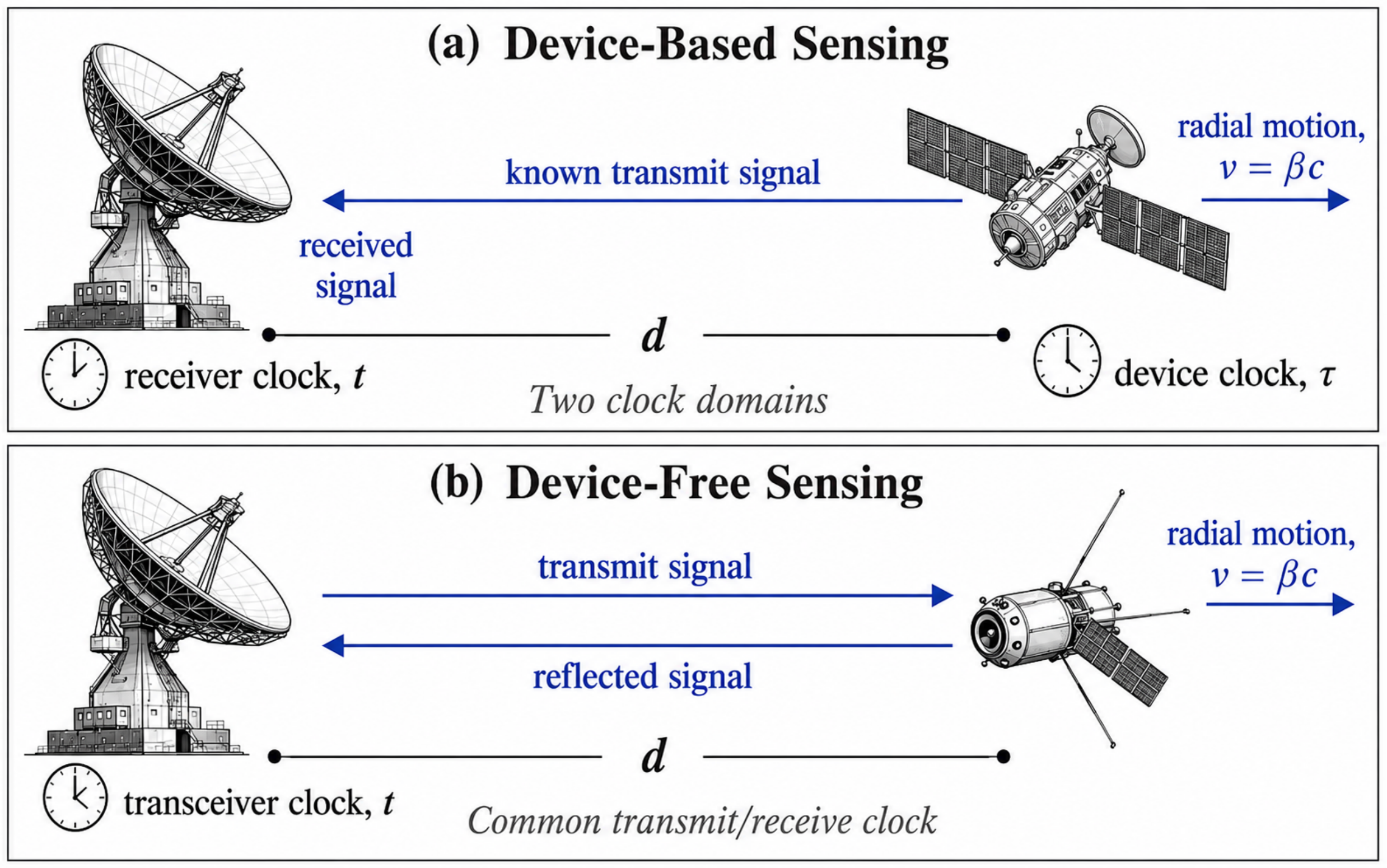}
	\caption{Illustration of the DB and DF sensing models in the receiver/transceiver rest frame.}
    \label{DB_DF_Fig}
\end{figure} 

\section{Relativistic Sensing Model}\label{sec:model}
\subsection{Device-Based Sensing}

Let us first consider a DB sensing scenario, where a moving device actively transmits a known signal to a receiver for range and velocity estimation. We take the receiver rest frame as the reference inertial frame, with $t$ denoting its \textit{coordinate time}. The device moves radially with a constant velocity $v=\beta c$, where $-1<\beta<1$ and $c$ is the speed of light. Let $\tau$ denote the \textit{proper time} measured by a clock co-moving with the device, and let $s(\tau)$ denote the known signal transmitted according to this proper time clock. We choose the time origins such that the reference sample $s(0)$ is emitted at $\tau=0$ and $t=0$, and denote the corresponding device-to-receiver range by $d>0$. The device trajectory in the receiver frame is therefore
\begin{equation}
r(t)=d+\beta ct,
\label{eq:db_trajectory}
\end{equation}
where $\beta>0$ and $\beta<0$ correspond to receding and approaching motion, respectively.

According to special relativity, the coordinate time and proper time along the device trajectory satisfy the differential relation ${\rm d}\tau={\rm d}t/\gamma$, where $\gamma = 1/\sqrt{1-\beta^2}$ is the \textit{Lorentz factor} \cite{Einstein1905}. Consider a signal sample emitted at receiver-frame coordinate time $t_e$ and device proper time $\tau_e$, and received at coordinate time $t_r$. We have $\tau_e=t_e/\gamma$. Since the electromagnetic wave propagates at speed $c$ in the receiver frame,
\begin{equation}
c(t_r-t_e)=r(t_e)=d+\beta ct_e.
\label{eq:db_propagation}
\end{equation}
It follows that
\begin{equation}
\tau_e=\sqrt{\frac{1-\beta}{1+\beta}}\left(t_r-\frac{d}{c}\right).
\label{eq:db_time_mapping}
\end{equation}
Hence, the received signal is
\begin{equation}
y_{\rm DB}(t_r)=\alpha s\left[\sqrt{\frac{1-\beta}{1+\beta}}\left(t_r-\frac{d}{c}\right)\right]+z(t_r),
\label{eq:db_signal}
\end{equation}
where $z(t)$ is the AWGN with variance of $\sigma^2$. We assume that the propagation coefficient $\alpha$ varies negligibly over the observation interval and can therefore be treated as a constant independent of $t_r$. This approximation remains accurate even at relativistic velocities when the waveform duration is sufficiently short relative to the propagation and geometric evolution time scales.

\subsection{Device-Free Sensing}

We next consider a monostatic DF sensing scenario, where a transceiver illuminates a moving target and receives its reflected signal. The transceiver rest frame is taken as the reference inertial frame. Suppose that it transmits a known signal $s(t)$. Consistent with the DB model, we define $d>0$ as the target-to-transceiver range when the moving target interacts with the reference sample $s(0)$, i.e., when $s(0)$ is scattered by the target. Since $s(0)$ is emitted at $t=0$, the corresponding scattering time is $d/c$, and the target trajectory is given as
\begin{equation}
r(t)=d+\beta c\left(t-\frac{d}{c}\right)
=(1-\beta)d+\beta ct.
\label{eq:df_trajectory}
\end{equation}

Consider a signal sample emitted at coordinate time $t_e$, scattered by the target at $t_s$, and received at $t_r$, where all times are measured in the transceiver frame. The forward and return propagation satisfy
\begin{equation}
c(t_s-t_e)=r(t_s),
\qquad
c(t_r-t_s)=r(t_s).
\label{eq:df_propagation}
\end{equation}
Using \eqref{eq:df_trajectory} and eliminating $t_s$ gives
\begin{equation}
t_e=\frac{1-\beta}{1+\beta}\left(t_r-\frac{2d}{c}\right).
\label{eq:df_time_mapping}
\end{equation}
Therefore, the received signal is
\begin{equation}
y_{\rm DF}(t_r)=\alpha s\left[\frac{1-\beta}{1+\beta}\left(t_r-\frac{2d}{c}\right)\right]+z(t_r).
\label{eq:df_signal}
\end{equation}
Unlike the DB model, no Lorentz transformation is involved, since both transmission and reception are referenced to the stationary transceiver clock. The same constant gain assumption are adopted as in the DB model.

\section{Fisher Information and Relativistic CRB}\label{sec:FIM}

\subsection{General Form of the Fisher Information Matrix}

Both the DB and DF signal models in Sec.~\ref{sec:model} can be written in the unified form
\begin{equation}
y(t)=\alpha s(at-b)+z(t),
\label{eq:unified_signal}
\end{equation}
where $a>0$ and $b$ are the time-scaling and time-shift parameters, respectively. Their dependence on range and velocity differs between the two sensing modes.

Write $\alpha=\alpha_{\rm R}+j\alpha_{\rm I}$ and define $\boldsymbol{\theta}:=[\alpha_{\rm R},\alpha_{\rm I},a,b]^T$. With $\mu(t):=\alpha s(at-b)$ denoting the noiseless received signal, the Fisher Information Matrix (FIM) under AWGN is
\begin{equation}
\left[\mathbf J_{\boldsymbol{\theta}}\right]_{mn}
=\frac{2}{\sigma^2}\operatorname{Re}\left\{
\int_{-\infty}^{\infty}
\frac{\partial\mu^*(t)}{\partial\theta_m}
\frac{\partial\mu(t)}{\partial\theta_n}{\rm d}t
\right\}.
\label{eq:general_fim}
\end{equation}
Define
\begin{equation}
g(t):=s(at-b),\quad
g_a(t):=t\dot{s}(at-b),\quad
g_b(t):=-\dot{s}(at-b),
\label{eq:signal_derivatives}
\end{equation}
where $\dot{s}(t):={\rm d}s(t)/{\rm d}t$. The corresponding derivatives are
\begin{equation}
\frac{\partial\mu}{\partial\alpha_{\rm R}}=g,\quad
\frac{\partial\mu}{\partial\alpha_{\rm I}}=jg,\quad
\frac{\partial\mu}{\partial a}=\alpha g_a,\quad
\frac{\partial\mu}{\partial b}=\alpha g_b.
\label{eq:mean_derivatives}
\end{equation}
Treating $\alpha$ as a nuisance parameter, the equivalent FIM (EFIM) for $(a,b)$ follows from the Schur complement as \cite{Jin1995,Das2020}
\begin{equation}
\left[\mathbf J_{a,b}\right]_{mn}
=
\rho\operatorname{Re}\left\{
\left\langle g_m,g_n\right\rangle
-\frac{\left\langle g_m,g\right\rangle\left\langle g,g_n\right\rangle}
{\left\|g\right\|^2}
\right\},
\label{eq:efim_ab}
\end{equation}
where $g_m,g_n\in\{g_a,g_b\}$, $\langle x,y\rangle:=\int_{-\infty}^{\infty}x^*(t)y(t){\rm d}t$, $\|x\|:=\sqrt{\langle x,x\rangle}$, and $\rho:=2|\alpha|^2/\sigma^2$. The Schur complement removes derivative components that are indistinguishable from a change in the unknown complex gain \cite{Stoica1989}.

Substituting \eqref{eq:signal_derivatives} into \eqref{eq:efim_ab} and changing the integration variable from $t$ to $at-b$ gives
\begin{align}
\label{eq:efim_entries}
&\left[\mathbf J_{a,b}\right]_{11}
\nonumber=\frac{\rho}{a^3}\left(\lambda_2+2b\lambda_1+b^2\lambda_0\right),\quad\left[\mathbf J_{a,b}\right]_{22}
=\frac{\rho}{a}\lambda_0,\\
&\left[\mathbf J_{a,b}\right]_{12} = \left[\mathbf J_{a,b}\right]_{21}
=-\frac{\rho}{a^2}\left(\lambda_1+b\lambda_0\right),
\end{align}
where the three waveform-dependent quantities are
\begin{align}
\label{eq:waveform_information_general}
&\nonumber\lambda_0:=\left\|\mathcal P_s^\perp\dot{s}\right\|^2,\;\;
\lambda_1:=\operatorname{Re}\left\{
\left\langle
\mathcal P_s^\perp(t\dot{s}),
\mathcal P_s^\perp\dot{s}
\right\rangle
\right\},\\
&\lambda_2:=\left\|\mathcal P_s^\perp(t\dot{s})\right\|^2,
\end{align}
and
\begin{equation}
\mathcal P_s^\perp f
:=
f-s\frac{\langle s,f\rangle}{\langle s,s\rangle}
\end{equation}
is the orthogonal projection onto the complement of the subspace spanned by $s$. Thus, $\lambda_0$ and $\lambda_2$ quantify the waveform information associated with time shift and time scaling, respectively, while $\lambda_1$ captures their coupling.

To relate these quantities to familiar time-frequency measures, consider a centered waveform
$s(t)=u(t)e^{j2\pi f_ct}$ with $S(f)$ being its Fourier transform, where $u(t)$ is real, even, and vanishes at infinity. Then $|s(t)|^2$ and $|\dot{s}(t)|^2$ are even in time, and $|S(f)|^2$ is symmetric about $f_c$. Define the signal energy, RMS duration, and RMS bandwidth as
\begin{equation}
E_s:=\int_{-\infty}^{\infty}|s(t)|^2{\rm d}t,\quad
T_{\rm rms}^2:=\frac{1}{E_s}\int_{-\infty}^{\infty}t^2|s(t)|^2{\rm d}t,
\label{eq:signal_time_measures}
\end{equation}
and
\begin{equation}
B_{\rm rms}^2
:=
\frac{1}{E_s}\int_{-\infty}^{\infty}
(f-f_c)^2|S(f)|^2{\rm d}f,
\label{eq:rms_bandwidth}
\end{equation}
Under these symmetry conditions,
\begin{equation}
\label{eq:waveform_information_symmetric}
\lambda_0=4\pi^2E_sB_{\rm rms}^2,\quad
\lambda_1=0,\quad
\lambda_2=\int_{-\infty}^{\infty}t^2|\dot{s}(t)|^2{\rm d}t-\frac{E_s}{4}.
\end{equation}
The EFIM then becomes
\begin{equation}
\mathbf J_{a,b}
=
\rho
\begin{bmatrix}
\dfrac{\lambda_2+b^2\lambda_0}{a^3}
&
-\dfrac{b\lambda_0}{a^2}
\\[6pt]
-\dfrac{b\lambda_0}{a^2}
&
\dfrac{\lambda_0}{a}
\end{bmatrix}.
\label{eq:closed_form_efim_ab}
\end{equation}
We next transform this EFIM to the physical parameters $(d,\beta)$ for the DB and DF sensing models.

\subsection{Device-Based CRB}

For the DB model in \eqref{eq:db_signal}, the affine parameters in \eqref{eq:unified_signal} are
\begin{equation}
a=\sqrt{\frac{1-\beta}{1+\beta}},
\qquad
b=\frac{ad}{c}.
\label{eq:db_ab}
\end{equation}
The corresponding Jacobian from $(d,\beta)$ to $(a,b)$ is
\begin{equation}
\mathbf G_{\rm DB}
:=
\frac{\partial(a,b)}{\partial(d,\beta)}
=
\begin{bmatrix}
0 & -\dfrac{a}{1-\beta^2}\\[8pt]
\dfrac{a}{c} & -\dfrac{ad}{c(1-\beta^2)}
\end{bmatrix}.
\label{eq:db_jacobian}
\end{equation}
Using \eqref{eq:closed_form_efim_ab}, the EFIM for $(d,\beta)$ is
\begin{equation}
\mathbf J_{d,\beta}^{\rm DB}
=
\mathbf G_{\rm DB}^{T}
\mathbf J_{a,b}
\mathbf G_{\rm DB}
=
\rho
\begin{bmatrix}
\dfrac{a\lambda_0}{c^2} & 0\\
0 & \dfrac{\lambda_2}{a(1-\beta^2)^2}
\end{bmatrix}.
\label{eq:db_fim}
\end{equation}
Thus, range and velocity are decoupled under the waveform symmetry conditions adopted above, and the reference range $d$ cancels out from the EFIM.

In the narrowband regime, where the envelope bandwidth is much smaller than $f_c$, we have $\dot{s}\left(t\right)\approx j2\pi f_c s\left(t\right)$ and $f_cT_{\rm rms}\gg 1$. The carrier contribution therefore dominates $\lambda_2$, yielding $\lambda_2\approx4\pi^2f_c^2E_sT_{\rm rms}^2$. Together with $\lambda_0=4\pi^2E_sB_{\rm rms}^2$, and $v=c\beta$, the corresponding CRBs are
\begin{align}
\varepsilon_{d}^{\rm DB}
&=
\frac{c^2}
{4\pi^2{\rm SNR}B_{\rm rms}^2}
\sqrt{\frac{1+\beta}{1-\beta}},
\label{eq:db_crb_d}\\
\varepsilon_{v}^{\rm DB}
&\approx
\frac{c^2}
{4\pi^2{\rm SNR}f_c^2T_{\rm rms}^2}
(1-\beta)^{\frac{5}{2}}(1+\beta)^{\frac{3}{2}},
\label{eq:db_crb_v}
\end{align}
where ${\rm SNR}:=\rho E_s$. Hence, the classical inverse dependence on the squared RMS bandwidth and RMS duration is preserved \cite{Liu2022FundamentalLimits}, while relativistic motion introduces additional velocity-dependent scaling factors. 


\subsection{Device-Free CRB}

For the DF model in \eqref{eq:df_signal}, the affine parameters are
\begin{equation}
a=\frac{1-\beta}{1+\beta},
\qquad
b=\frac{2ad}{c}.
\label{eq:df_ab}
\end{equation}
The corresponding Jacobian is
\begin{equation}
\mathbf G_{\rm DF}
:=
\frac{\partial(a,b)}{\partial(d,\beta)}
=
\begin{bmatrix}
0 & -\dfrac{2}{(1+\beta)^2}\\[8pt]
\dfrac{2a}{c} & -\dfrac{4d}{c(1+\beta)^2}
\end{bmatrix}.
\label{eq:df_jacobian}
\end{equation}
The EFIM for $(d,\beta)$ is
\begin{equation}
\mathbf J_{d,\beta}^{\rm DF}
=
\mathbf G_{\rm DF}^{T}
\mathbf J_{a,b}
\mathbf G_{\rm DF}
=
4\rho
\begin{bmatrix}
\dfrac{a\lambda_0}{c^2} & 0\\
0 & \dfrac{\lambda_2}{a(1-\beta^2)^2}
\end{bmatrix}.
\label{eq:df_fim}
\end{equation}
As in the DB case, range and velocity are decoupled and the reference range $d$ cancels out. The DF model has the same EFIM structure but a different relativistic scaling factor $a$ and an additional factor of 4 due to the two-way propagation. Using the same waveform relations, the DF CRBs become
\begin{align}
\varepsilon_{d}^{\rm DF}
&=
\frac{c^2}
{16\pi^2{\rm SNR}B_{\rm rms}^2}
\frac{1+\beta}{1-\beta},
\label{eq:df_crb_d}\\
\varepsilon_{v}^{\rm DF}
&\approx
\frac{c^2}
{16\pi^2{\rm SNR}f_c^2T_{\rm rms}^2}
(1-\beta)^3(1+\beta).
\label{eq:df_crb_v}
\end{align}

\section{Asymptotic Analysis}\label{sec:asym}
\subsection{Low-Speed Limits}

We first consider the low-speed regime $|\beta|\ll1$. As $\beta\rightarrow0$, both time-scaling factors approach 1, and the relativistic CRBs recover their classical low-speed forms. For DB sensing,
\begin{equation}
\varepsilon_{d}^{\rm DB}
\rightarrow
\frac{c^2}{4\pi^2{\rm SNR}B_{\rm rms}^2},
\qquad
\varepsilon_{v}^{\rm DB}
\rightarrow
\frac{c^2}{4\pi^2{\rm SNR}f_c^2T_{\rm rms}^2},
\label{eq:db_low_speed}
\end{equation}
whereas for DF sensing,
\begin{equation}
\varepsilon_{d}^{\rm DF}
\rightarrow
\frac{c^2}{16\pi^2{\rm SNR}B_{\rm rms}^2},
\qquad
\varepsilon_{v}^{\rm DF}
\rightarrow
\frac{c^2}{16\pi^2{\rm SNR}f_c^2T_{\rm rms}^2}.
\label{eq:df_low_speed}
\end{equation}
Thus, the classical inverse dependence on RMS bandwidth and RMS duration is recovered, while the two-way propagation in DF sensing yields a reduction in both CRBs relative to DB sensing with a factor of 4.

\subsection{Ultrarelativistic Limits}

We next examine the ultrarelativistic limits $\beta\rightarrow1$ and $\beta\rightarrow-1$, corresponding to receding and approaching motion at speeds arbitrarily close to $c$, respectively. For a consistent comparison, all CRBs are normalized by the corresponding DB CRB at $\beta=0$.

For a receding object with $\beta\rightarrow1$, the normalized CRBs satisfy
\begin{align}
\nonumber\frac{\varepsilon_{d}^{\rm DB}(\beta)}
{\varepsilon_{d}^{\rm DB}(0)}
&\sim
\sqrt{\frac{2}{1-\beta}},
&
\frac{\varepsilon_{d}^{\rm DF}(\beta)}
{\varepsilon_{d}^{\rm DB}(0)}
&\sim
\frac{1}{2(1-\beta)},\\
\frac{\varepsilon_{v}^{\rm DB}(\beta)}
{\varepsilon_{v}^{\rm DB}(0)}
&\sim
2^{\frac{3}{2}}(1-\beta)^{\frac{5}{2}},
&
\frac{\varepsilon_{v}^{\rm DF}(\beta)}
{\varepsilon_{v}^{\rm DB}(0)}
&\sim
\frac{1}{2}(1-\beta)^3.
\label{eq:receding_limit}
\end{align}
Hence, the range CRBs of both sensing modes diverge as $\beta\rightarrow1$, with the DF CRB increasing faster than the DB CRB. In contrast, both velocity CRBs approach zero, with the DF CRB decreasing faster.

For an approaching object with $\beta\rightarrow-1$, we have
\begin{align}
\nonumber\frac{\varepsilon_{d}^{\rm DB}(\beta)}
{\varepsilon_{d}^{\rm DB}(0)}
&\sim
\sqrt{\frac{1+\beta}{2}},
&
\frac{\varepsilon_{d}^{\rm DF}(\beta)}
{\varepsilon_{d}^{\rm DB}(0)}
&\sim
\frac{1+\beta}{8},\\
\frac{\varepsilon_{v}^{\rm DB}(\beta)}
{\varepsilon_{v}^{\rm DB}(0)}
&\sim
2^{\frac{5}{2}}(1+\beta)^{\frac{3}{2}},
&
\frac{\varepsilon_{v}^{\rm DF}(\beta)}
{\varepsilon_{v}^{\rm DB}(0)}
&\sim
2(1+\beta).
\label{eq:approaching_limit}
\end{align}
Thus, both range and velocity CRBs approach zero as $\beta\rightarrow-1$. The DF range CRB decreases faster than the DB range CRB, whereas the DB velocity CRB decreases faster than its DF counterpart. These different asymptotic orders result from the distinct one-way and two-way time-scaling laws of DB and DF sensing.

\section{Numerical Results}\label{sec:numerical}

We numerically illustrate the relativistic CRB scaling laws derived above. The DB and DF CRBs are normalized by the corresponding DB values at $\beta=0$, so the resulting curves depend only on $\beta$ and the sensing mode. Since practical DF sensing may suffer additional two-way propagation loss and target reflection loss, we assume equal received SNR for DB and DF sensing to separate these link-budget effects from the relativistic kinematic effects of interest.

\begin{figure}[!t]
	\centering
	\includegraphics[width = \columnwidth]{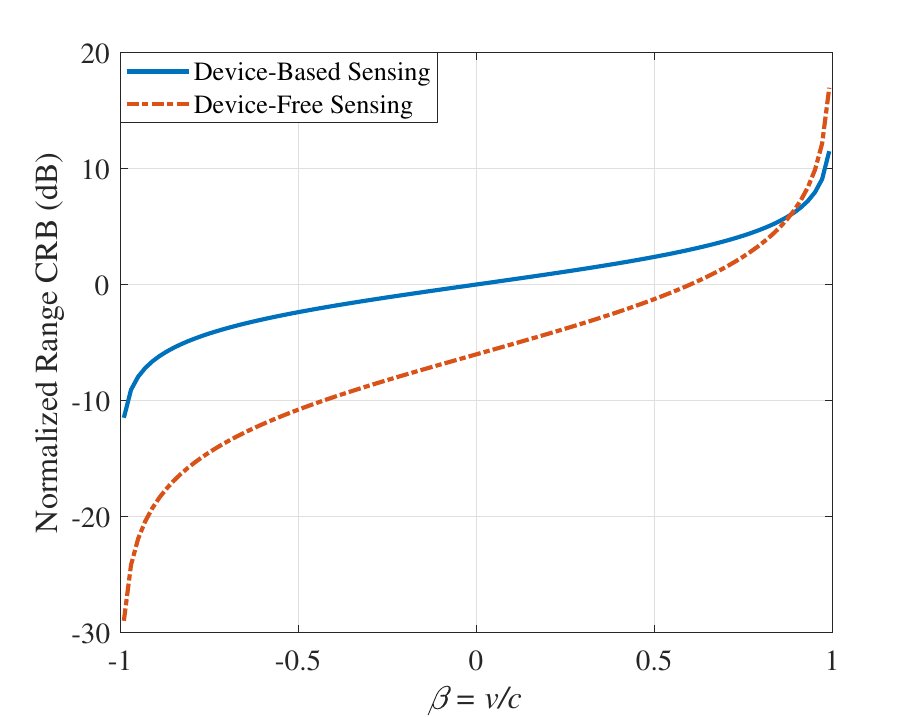}
	\caption{Normalized range CRBs.}
    \label{range_crb_fig}
\end{figure} 

\begin{figure}[!t]
	\centering
	\includegraphics[width = \columnwidth]{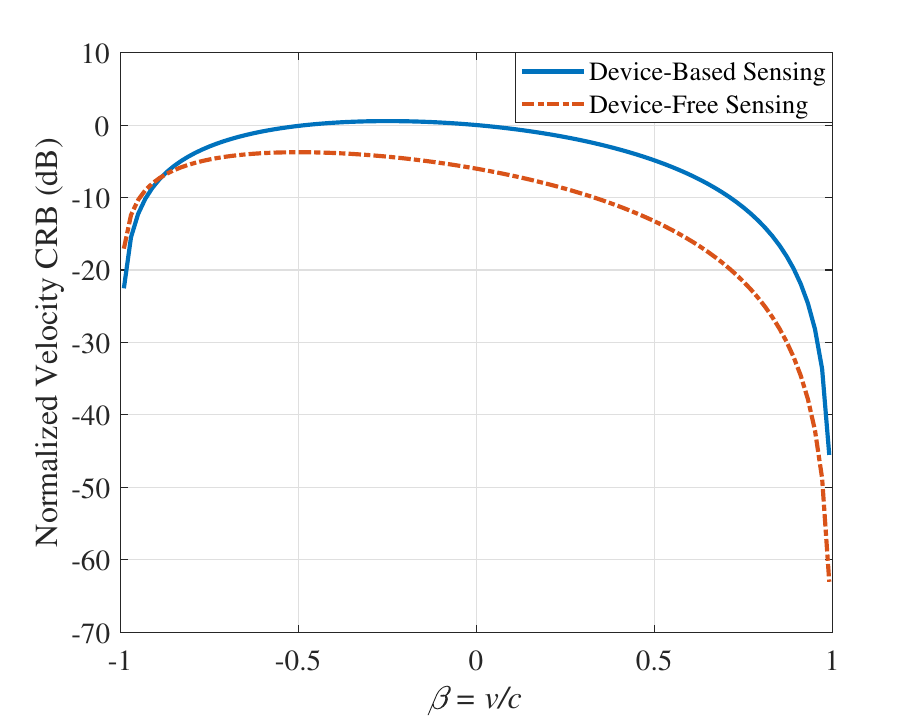}
	\caption{Normalized velocity CRBs.}
    \label{vel_crb_fig}
\end{figure} 

Fig.~\ref{range_crb_fig} shows the normalized range CRBs. At $\beta=0$, the DF CRB is $6$ dB lower than the DB CRB under equal received SNR, reflecting the higher range sensitivity associated with two-way propagation. As $\beta$ increases, both CRBs increase because the received signal is increasingly stretched and the observable time shift becomes less sensitive to the reference range. This loss of sensitivity is stronger for DF sensing, so its CRB grows faster and eventually exceeds the DB CRB. As $\beta\rightarrow1$, both range CRBs diverge and the Fisher information for $d$ vanishes, so the range becomes unidentifiable in this limit. For an approaching object, waveform compression produces the opposite effect: changes in range induce more pronounced changes in the received signal, so both CRBs decrease and eventually tend to zero as $\beta\rightarrow-1$, with a faster decay for DF sensing. These opposite trends reflect the asymmetric effect of relativistic time scaling on receding and approaching motion.

Fig.~\ref{vel_crb_fig} shows the normalized velocity CRBs. At $\beta=0$, the DF CRB is again $6$ dB lower than the DB CRB under equal received SNR. Unlike the range CRB, both velocity CRBs decrease toward zero as $\beta$ approaches either $1$ or $-1$. This behavior originates from the relativistic time-scaling law. As $|\beta|$ approaches 1, the time-scaling factor becomes increasingly sensitive to velocity, so even a small change in $v$ produces a pronounced compression or stretching of the received signal. The Fisher information for velocity therefore grows without bound and the corresponding CRB tends to zero. The different decay rates of DB and DF sensing arise from their different one-way and two-way time-scaling laws, with DF decreasing faster as $\beta\rightarrow1$ and DB decreasing faster as $\beta\rightarrow-1$. Taken together, the two figures show that receding motion suppresses range information while enhancing velocity information, whereas approaching motion drives both range and velocity CRBs toward zero.

\section{Conclusion}\label{sec:conc}

This letter studied range and velocity estimation for device-based and device-free sensing under relativistic motion. By expressing both sensing modes through a common time-scaling and time-shift model, we derived closed-form CRBs and characterized their dependence on waveform bandwidth, duration, and object velocity. The results recover the classical low-speed CRBs while revealing distinct ultrarelativistic behaviors for approaching and receding motion, with different scaling laws for DB and DF sensing. These results provide a simple estimation-theoretic characterization of relativistic sensing. Extension to multi-dimensional motion and bistatic/multi-static sensing is left for future work.
    \balance
	\bibliographystyle{IEEEtran}
	\bibliography{relativistic_CRB_bib}

\begin{thebibliography}{10}
\providecommand{\url}[1]{#1}
\csname url@samestyle\endcsname
\providecommand{\newblock}{\relax}
\providecommand{\bibinfo}[2]{#2}
\providecommand{\BIBentrySTDinterwordspacing}{\spaceskip=0pt\relax}
\providecommand{\BIBentryALTinterwordstretchfactor}{4}
\providecommand{\BIBentryALTinterwordspacing}{\spaceskip=\fontdimen2\font plus
\BIBentryALTinterwordstretchfactor\fontdimen3\font minus
  \fontdimen4\font\relax}
\providecommand{\BIBforeignlanguage}[2]{{%
\expandafter\ifx\csname l@#1\endcsname\relax
\typeout{** WARNING: IEEEtran.bst: No hyphenation pattern has been}%
\typeout{** loaded for the language `#1'. Using the pattern for}%
\typeout{** the default language instead.}%
\else
\language=\csname l@#1\endcsname
\fi
#2}}
\providecommand{\BIBdecl}{\relax}
\BIBdecl

\bibitem{Einstein1905}
A.~Einstein, ``Zur elektrodynamik bewegter k{\"o}rper,'' \emph{Ann. Phys.},
  vol. 322, no.~10, pp. 891--921, 1905.

\bibitem{Friedlander1984}
B.~Friedlander, ``On the {Cram\'er--Rao} bound for time delay and {D}oppler
  estimation,'' \emph{IEEE Trans. Inf. Theory}, vol.~30, no.~3, pp. 575--580,
  1984.

\bibitem{Dogandzic2001}
A.~Dogandzic and A.~Nehorai, ``{Cram\'er--Rao} bounds for estimating range,
  velocity, and direction with an active array,'' \emph{IEEE Trans. Signal
  Process.}, vol.~49, no.~6, pp. 1122--1137, 2001.

\bibitem{Lindegren2003}
L.~Lindegren and D.~Dravins, ``The fundamental definition of ``radial
  velocity'','' \emph{Astron. Astrophys.}, vol. 401, no.~3, pp. 1185--1201,
  2003.

\bibitem{Wright2014}
J.~T. Wright and J.~D. Eastman, ``Barycentric corrections at 1 cm s$^{-1}$ for
  precise {D}oppler velocities,'' \emph{Publ. Astron. Soc. Pac.}, vol. 126, no.
  943, pp. 838--852, 2014.

\bibitem{Harkins1979}
M.~D. Harkins, ``The relativistic {D}oppler shift in satellite tracking,''
  \emph{Radio Sci.}, vol.~14, no.~4, pp. 671--675, 1979.

\bibitem{Turyshev2013}
S.~G. Turyshev, V.~T. Toth, and M.~V. Sazhin, ``General relativistic
  observables of the {GRAIL} mission,'' \emph{Phys. Rev. D}, vol.~87, no.~2, p.
  024020, 2013.

\bibitem{Hees2014}
A.~Hees, S.~Bertone, and C.~L. Poncin-Lafitte, ``Relativistic formulation of
  coordinate light time, {D}oppler, and astrometric observables up to the
  second post-{M}inkowskian order,'' \emph{Phys. Rev. D}, vol.~89, no.~6, p.
  064045, 2014.

\bibitem{deHoop2009}
A.~T. de~Hoop, ``Electromagnetic radiation from moving, pulsed source
  distributions: {T}he 3{D} time-domain relativistic {D}oppler effect,''
  \emph{Wave Motion}, vol.~46, no.~1, pp. 74--77, 2009.

\bibitem{Kelly1965}
E.~J. Kelly and R.~P. Wishner, ``Matched-filter theory for high-velocity,
  accelerating targets,'' \emph{IEEE Trans. Mil. Electron.}, vol.~9, no.~1, pp.
  56--69, 1965.

\bibitem{Rihaczek1967}
A.~W. Rihaczek, ``Delay-{D}oppler ambiguity function for wideband signals,''
  \emph{IEEE Trans. Aerosp. Electron. Syst.}, vol. AES-3, no.~4, pp. 705--711,
  1967.

\bibitem{Altes1973}
R.~A. Altes, ``Some invariance properties of the wide-band ambiguity
  function,'' \emph{J. Acoust. Soc. Am.}, vol.~53, no.~4, pp. 1154--1160, 1973.

\bibitem{Sibul1981}
L.~H. Sibul and E.~L. Titlebaum, ``Volume properties for the wideband ambiguity
  function,'' \emph{IEEE Trans. Aerosp. Electron. Syst.}, vol. AES-17, no.~1,
  pp. 83--87, 1981.

\bibitem{White1992}
L.~B. White, ``The wide-band ambiguity function and altes' {Q}-distribution:
  Constrained synthesis and time-scale filtering,'' \emph{IEEE Trans. Inf.
  Theory}, vol.~38, no.~2, pp. 886--892, 1992.

\bibitem{Weiss1994}
L.~G. Weiss, ``Wavelets and wideband correlation processing,'' \emph{IEEE
  Signal Process. Mag.}, vol.~11, no.~1, pp. 13--32, 1994.

\bibitem{Borden2006}
B.~Borden, ``On the fractional wideband and narrowband ambiguity function in
  radar and sonar,'' \emph{IEEE Signal Process. Lett.}, vol.~13, no.~9, pp.
  545--548, 2006.

\bibitem{Jin1995}
Q.~Jin, K.~M. Wong, and Z.-Q. Luo, ``The estimation of time delay and {D}oppler
  stretch of wideband signals,'' \emph{IEEE Trans. Signal Process.}, vol.~43,
  no.~4, pp. 904--916, 1995.

\bibitem{Giunta1998}
G.~Giunta, ``Fast estimators of time delay and {D}oppler stretch based on
  discrete-time methods,'' \emph{IEEE Trans. Signal Process.}, vol.~46, no.~7,
  pp. 1785--1797, 1998.

\bibitem{Niu1999}
X.~X. Niu, P.~C. Ching, and Y.~T. Chan, ``Wavelet based approach for joint time
  delay and {D}oppler stretch measurements,'' \emph{IEEE Trans. Aerosp.
  Electron. Syst.}, vol.~35, no.~3, pp. 1111--1119, 1999.

\bibitem{Das2020}
P.~Das, J.~Vil{\`a}-Valls, F.~Vincent, L.~Davain, and E.~Chaumette, ``A new
  compact delay, {D}oppler stretch and phase estimation {CRB} with a
  band-limited signal for generic remote sensing applications,'' \emph{Remote
  Sens.}, vol.~12, no.~18, p. 2913, 2020.

\bibitem{Liu2022FundamentalLimits}
A.~Liu, Z.~Huang, M.~Li, Y.~Wan, W.~Li, T.~X. Han, C.~Liu, R.~Du, D.~K.~P. Tan,
  J.~Lu, Y.~Shen, F.~Colone, and K.~Chetty, ``A survey on fundamental limits of
  integrated sensing and communication,'' \emph{IEEE Commun. Surveys Tuts.},
  vol.~24, no.~2, pp. 994--1034, 2022.

\bibitem{Stoica1989}
P.~Stoica and A.~Nehorai, ``{MUSIC}, maximum likelihood, and {Cram\'er--Rao}
  bound,'' \emph{IEEE Trans. Acoust., Speech, Signal Process.}, vol.~37, no.~5,
  pp. 720--741, 1989.

\end{thebibliography}

\end{document}